# From (Elementary) Mathematical Data Model Schemas to Safe Blazor Web Applications with Claude AI

**Christian Mancas\* and Diana Christina Mancas**

*Mathematics and Computer Science Department, Ovidius University at Constanta, Romania*

**\*Corresponding Author:** Christian Mancas, Ovidius University, Bd. Mamaia 124, Constanta, CT, Romania, Email: christian.mancas@gmail.com

## Abstract

This research paper mainly describes how to develop MS Blazor safe web applications with Claude AI, version Sonnet 4.5, starting from (Elementary) Mathematical Data Model schemas. In the sequel, it also provides a list of general software engineering best practice rules, as well as issues of the MS Blazor Server platform.

## Introduction

Half a century ago, as a 4th grade student in Computers at the Bucharest Polytechnic University, during a course of Computer Architectures, the corresponding author of this paper asked one of their Professors, lecturing on computer technologies advances, when would start humanity talk mathematics to computers, and was very disappointed by his "Never!" answer. Ten years later, he published his first paper on a(n) (Elementary) Mathematical Data Model ((E)MDM) (see [1]). Moreover, he also started to develop *MatBase*[2], an intelligent data and knowledge base management system prototype based mainly on (E)MDM, but also on Datalog¬ [3], the Relational (RDM) [3–5] and Entity-Relationship (E-R) [5–7] Data Models.

*MatBase* was designed to accept (E)MDM schemas (which include Datalog¬ programs), translate them to relational databases (rdbs) and automatically code generated database (db) applications (apps), to accept and import rdbs and translate them to (E)MDM schemas, to accept E-R Diagrams and translate them to (E)MDMD schemas, as well as, again dually, to translate (E)MDM schemas into E-R diagrams [8]. The main goal was always to reach modeling as programming [9] and, especially, mathematical data modeling.

During all following years up until today, both (E)MDM and *MatBase* ever evolved, the former by incorporating new constraint types, the latter especially trying to keep up with newer software technologies. Since many years already, *MatBase* has two versions, one for small and medium dbs developed in VBA over MS Access, and one for large dbs developed in C# over MS SQL Server.

Our latest project was to develop a *MatBase* safe web version, based on MS Blazor and Azure / SQL Server. With the advent of Artificial Intelligence (AI) Large Language Models (LLMs) [10], we decided, as most of today's developers, to benefit from AI coding platforms, to significantly improve our productivity. Consequently, we started to consider all top 5 AI programming tools this year: GitHub [11], Cursor [12], Windsurf [13], Replit [14], and Claude AI [15]. All of them are referred to as AI Coding Agents, Code Assistants, or AI IDEs (Integrated Development Environments) providing "*vibe coding*", i.e., the ability to create software simply by describing it.

We opted for Claude AI (which is integrated with GitHub and Claude Code [16–17]) for two main reasons: accessibility, as it is also available directly from any internet browser (i.e., it's terminal-based, which makes integrating its agents easier into much wider ecosystems than the IDE ones), and reputation, as it is used by the Pentagon, Amazon, Microsoft, and 300,000+ companies around the world.

Claude Code, released only one year ago, on February 24, 2025, by the U.S. company Anthropic, as one of their Claude AI's tools, is a *supervised coding agent*, i.e., a coding assistant able of performing autonomously even sophisticated software engineering tasks.

Claude Code supports most of the currently used programming languages, from SQL, C++, Java, C#, Python, Go, to JavaScript and TypeScript, from HTML, XML, and shell scripts (e.g., Windows PowerShell), to JSON, YAML, and TOML, but also legacy ones like VBA or PHP, and even Datalog, so, very probably, Prolog as well. For all these languages, it excels from syntax recognition, idioms, ecosystem conventions, and best practices knowledge to deep understanding of language-specific patterns. This allows it to reverse-engineer, refactor, and generate code for all languages it supports, as well as to provide intelligent hints aligned with corresponding developer community standards.

Moreover, Claude Code has extensive knowledge of complex frameworks like Django, React, Express.js, Razor, Blazor, etc., as well as package management systems (e.g., npm, pip, cargo, etc.), and excels in dealing with modern polyglot development projects, using, e.g., C#ASP.NET, JavaScript, HTML, CSS, C#, and different flavors of embedded or/and extended SQL, all within a single project repository.

As another plus, it is reported [18] that you may use up to 15 parallel Claude Code sessions working on the same project. We used 2 of them.

To get acquainted with Claude AI, we decided to first build from scratch a medium-sized complex safe web app using its coding capabilities, choosing to target a local MS Windows Service developed with the MS Blazor and SQL Servers technologies.

Simpler than the Model-View-Controller architecture, the file-based, page-focused MS *Razor* one creates safe dynamic web content through a markup syntax used in ASP.NET Core, embedding C# code directly into HTML. This enables developers to build fast and secure web applications made of *Razor Pages*, also isolating the CSS code.

MS *Blazor Server* is a free, open-source web framework that allows developers to create interactive, component-based web graphical user interfaces (GUIs), using only C# and HTML (instead of JavaScript) in Razor pages. Blazor enables full-stack development completely within the .NET ecosystem, for building secure, high-performance web apps that may run both in browsers, via WebAssembly, as well as on web servers.

A web db application is *safe* (*secure*) [19] if, according to the Open Web Application Security Project (OWASP), it is guarded at least against the following Top 10 Web Application Vulnerabilities in 2025:

1. Broken Access Control (Present in nearly 1 / 25 applications tested by OWASP.)
2. Security Misconfiguration (Increasing risk with the shift towards highly configurable software.)
3. Software Supply Chain Failures (Use of obsolete function or unmaintained third-party components, reliance on not updateable components, and dependency on vulnerable third-party ones.)
4. Cryptographic Failures (Root cause of sensitive data exposure.)
5. Injection (Attackers' malicious code injection into SQL statements.)
6. Insecure Design (Poor control design, e.g., exposed sensitive data, like connection strings to db servers.)
7. Identification and Authentication Failures (A common vector for attacks.)
8. Software or Data Integrity Failures (Code or infrastructure introduced without checking for integrity.)
9. Security Logging and Monitoring Failures (Difficult to test for, but key for detecting breaches.)
10. Mishandling of Exceptional Conditions (Improper error handling, logical errors, failing open, and other related scenarios stemming from abnormal conditions that systems may encounter.)

The next Sections explore related work, present the materials and methods used, the results obtained, a discussion, conclusions, and the reference list.

**Related Work**

Details on the supplementary Opus Claude AI version features are given in [20]. Guidelines for developing with Claude Code are presented in [21–24]. A coding capabilities comparison between ChatGPT [25] and Claude is provided in [26]. An analysis on why Claude Code is becoming the go-to alternative to expensive coding assistants is the main topic of [27]. The current Top 5 AI app builders are available through [28]. A less fortunate experience with Claude Code is discussed in [29]. Probably the best reference for Claude Code capabilities is the podcast of Boris Cherny, the inventor and current Head of Claude Code at Anthropic [30].

**Materials and Methods**

We used Claude Code's free Sonnet 4.5 version. Our platforms were 2 Core i7 Toshiba Satellite notebooks, running Windows 10, SQL Server 21, and .NET 8.

We did not follow any guidance in interacting with Claude AI: we were very curious to see whether it understands (E)MDM schemas, and we were extremely pleasantly surprised that it does; then, for any GUI requests, such as layout, menus, submenus, filters, etc., we only used plain English, just like between developers. That is pure vibe-coding flavored with function, first-order predicate logic formula, and Datalog rule definitions.

For example, our dialogue with Claude AI for this project started as shown in Figure 1, where, besides the implicit codomain and totality (not null) constraints, there are 7 explicit ones, formalizing the following "business rules": ages of male and female persons must be integers at most equal to 140; nobody may be his/her/its own ancestor or descendant; mothers must have sex 'F'; fathers must have sex 'M'; neutral persons (e.g., occupying armies, international bodies, triumvirates, etc.) may not have parents; mothers may give birth only when between 5 and 75, but not after death; fathers may have children only between 9 and 100, but only at most 1 year after death.

In less than 15 minutes, Claude AI gave us the answer shown in Figures 2 and 3, and generated all 16 files needed for creating the corresponding db and web app. One of them was a README file, with details on the implementation as shown in Figures 4 to 7.

After inspecting the code, we asked Claude to reverse the 3 renaming it did (namely, *id* for *x*, *MotherId* for *Mother*, and *FatherId* for *Father*), to implement *CurrentYear*() as YEAR(GETDATE()) instead of its 9999 (which would make redundant two check constraints), add *Age* to table *PERSONS* as a computed column, not as a view one, and use YEAR(GETDATE()) in its expression as well. Claude AI promptly changed the generated code accordingly.

We were very pleasantly surprised by the fact that it also added a commonsense constraint not present in our request: $(\forall x \in PERSONS)(BirthYear(x) \leq PassedAwayYear(x))$.

We ran the provided scripts according to its setup instructions (see Figure 7) and the first iteration of our safe web app was immediately up and running! Then, we asked Claude to enhance the app from a Razor to a Blazor Server one. It rapidly provided us with corresponding new versions of the app files, and we successfully rebuild the app. This change allowed for enforcing the constraints (that Claude calls *axioms*) on *Father* and *Mother* in the GUI, in C#, without JavaScript and, more importantly, avoid sending unplausible data for them to the SQL Server.

For deployment, Claude AI proposed two solutions: the old-fashioned Windows Server with IIS or the modern Azure App Service + Azure SQL one. Being a test, we opted for a third one: a local Windows service + our local SQL Server. It provided us immediately with the needed PowerShell script and code updates, and, after correcting them 12 times (e.g., it forgot to grant access rights for the db to the Windows service), according to the error messages issued by the builder or the deployer, as well as inspecting the messages in the Console of the browser's ChromeDevTools, our web app was ready for testing.

Hi Claude!

How would you create a safe Razor Pages web app using C#ASP.NET and SQL Server for the following math scheme:

$x : \text{PERSONS} \leftrightarrow \text{NAT}(16)$

$\text{Name} : \text{PERSONS} \rightarrow \text{UNICODE}(128)$

$\text{Sex} : \text{PERSONS} \rightarrow \{\text{"F"}, \text{"M"}, \text{"N"}\}$

$\text{Mother} : \text{PERSONS} \rightarrow \text{PERSONS} \cup \text{NULLS}$

$\text{Father} : \text{PERSONS} \rightarrow \text{PERSONS} \cup \text{NULLS}$

$\text{BirthYear} : \text{PERSONS} \rightarrow [-6000, \text{CurrentYear}()] \cup \text{NULLS}$

$\text{PassedAwayYear} : \text{PERSONS} \rightarrow [-6000, \text{CurrentYear}()] \cup \text{NULLS}$

$\text{Age} : \text{PERSONS} \rightarrow [0, 140] \cup \text{NULLS}$

$(\forall x \in \text{PERSONS})(\text{Age}(x) = \text{isNull}(\text{PassedAwayYear}(x), \text{CurrentYear}()) - \text{BirthYear}(x))$

$(\forall x \in \text{PERSONS})(\text{Sex}(x) \neq \text{'N'} \Rightarrow 0 \leq \text{Age}(x) \leq 140)$

Mother • Father acyclic

$(\forall x \in \text{PERSONS})(\text{Sex}(\text{Mother}(x)) = \text{'F'})$

$(\forall x \in \text{PERSONS})(\text{Sex}(\text{Father}(x)) = \text{'M'})$

$(\forall x \in \text{PERSONS})(\text{Sex}(x) = \text{'N'} \Rightarrow \text{Mother}(x) \in \text{NULLS} \wedge \text{Father}(x) \in \text{NULLS})$

$(\forall x \in \text{PERSONS})(5 \leq \text{BirthYear}(x) - \text{BirthYear}(\text{Mother}(x)) \leq 75 \wedge \text{BirthYear}(x) \leq \text{PassedAwayYear}(\text{Mother}(x)))$

$(\forall x \in \text{PERSONS})(9 \leq \text{BirthYear}(x) - \text{BirthYear}(\text{Father}(x)) \leq 100 \wedge \text{BirthYear}(x) \leq \text{PassedAwayYear}(\text{Father}(x)) + 1)$

*Figure 1*: Start of our dialogue with Claude AI

I'll analyze this math scheme carefully and create a comprehensive Razor Pages web app. Let me first read the frontend design skill.

Reading frontend design skill

Now let me carefully decode the math scheme and create the full application:

Created 22 files, ran 3 commands, read a file

Here's a complete walkthrough of every decision made:

Formal Spec → Implementation

$x : \text{PERSONS} \leftrightarrow \text{NAT}(16)$ — NAT(16) means a 16-bit natural number, so Id ∈ [0, 65535]. Stored as BIGINT in SQL (not SMALLINT, which is signed) with CHECK (Id BETWEEN 0 AND 65535), and validated in C# before the DB round-trip.

Name : PERSONS → UNICODE(128) — NVARCHAR(128) NOT NULL. The N prefix is essential — it's the SQL Server type that stores actual Unicode (UCS-2/UTF-16), not just ASCII.

Sex : PERSONS → {"F","M","N"} — CHAR(1) with a CHECK constraint. The UI dropdowns only show valid values, but the constraint enforces correctness at the DB level regardless of how data arrives.

Age (derived) — Never stored. Computed in a VIEW as ISNULL(PassedAwayYear, YEAR(GETDATE())) – BirthYear, exactly matching Axiom 1.

*Figure 2*: First Claude AI answer (first half)

What was equally impressive was the fact that, for every error, Claude explains to you the reasons why it occurred, not only the solution to correct it. We also like the fact that the footer of any dialog page with Claude reads "Claude is AI and can make mistakes. Please double-check responses."

Encouraged by this unexpected success, we continued to fetch it, in 3 subsequent steps, the (E)MDM schemas shown in Figures 8 to 12. The explicit constraints from Figure 8 correspond to the following 7 "business rules": no parent gives the same name to two of her/his children; no spouses couple get marry or divorce several times in the same year; no marriage may end before beginning; wives' sex must be 'F'; husbands' sex must be 'M'. The 6 ones from Figure 9 correspond to: nobody may get married / divorce before being born or after death; incestuous marriages (i.e., between siblings or parent and child) are forbidden; marriage years must be naturals at most equal to 140; no spouse may remarry while being married; for any marriage, husband and wife must have been simultaneously alive at least one day of the same year.

**All 8 Axioms — Where Each Is Enforced**

Every axiom is enforced at **two independent layers** (defence in depth):

| # | Axiom | C# layer | SQL layer |
|---|---|---|---|
| ① | Age formula | Person.ComputeAge() | View definition |
| ② | Sex≠'N' ⟹ Age∈[0,140] | ValidateStandalone | Trigger |
| ③ | Mother·Father acyclic | self-ref check | FN_HasAncestorCycle CTE in trigger |
| ④ | Sex(Mother)='F' | ValidateWithParents | Trigger + FK |
| ⑤ | Sex(Father)='M' | ValidateWithParents | Trigger + FK |
| ⑥ | Sex='N' ⟹ no parents | ValidateStandalone | CHECK constraint |
| ⑦ | Mother gap 5–75 yrs + alive | ValidateWithParents | Trigger |
| ⑧ | Father gap 9–100 yrs + posthumous+1 | ValidateWithParents | Trigger |

---

**Security Measures**

- **SQL Injection** — zero ad-hoc SQL; all DB calls go through named stored procedures via Dapper with typed parameters
- **CSRF** — anti-forgery tokens on every POST, enforced globally
- **XSS** — Razor's automatic HTML-encoding; strict Content-Security-Policy header
- **Clickjacking** — X-Frame-Options: DENY
- **HTTPS** — UseHttpsRedirection() + HSTS (365 days)
- **Over-posting** — [BindProperty] on a dedicated Input model; Id is readonly on the edit form

*Figure 3*: First Claude AI answer (second half)

---

**Persons Registry — ASP.NET Core Razor Pages + SQL Server**

A safe, formally correct web application implementing the **PERSONS** mathematical scheme.

---

**Formal Specification → Implementation Mapping**

| Formal element | C# / SQL implementation |
|---|---|
| x : PERSONS ≪ NAT(16) | Id BIGINT with CHECK (Id BETWEEN 0 AND 65535) + C# range validation |
| Name : PERSONS → UNICODE(128) | Name NVARCHAR(128) NOT NULL |
| Sex : PERSONS → {"F","M","N"} | Sex CHAR(1) with CHECK (Sex IN ('F','M','N')) |
| Mother : PERSONS → PERSONS ∪ NULLS | MotherId BIGINT NULL REFERENCES Persons(Id) |
| Father : PERSONS → PERSONS ∪ NULLS | FatherId BIGINT NULL REFERENCES Persons(Id) |
| BirthYear : PERSONS → [-6000, CurrentYear()] ∪ NULLS | BirthYear SMALLINT NULL + CHECK + trigger |
| PassedAwayYear : PERSONS → [-6000, CurrentYear()] ∪ NULLS | PassedAwayYear SMALLINT NULL + CHECK + trigger |
| Age : PERSONS → [0,140] ∪ NULLS | Computed column in PersonsWithAge VIEW |

*Figure 4*: README file (first part)

**Layer 2 — SQL Server (Schema.sql) — database-level guarantee**

- CHECK constraints for domain ranges and Sex
- FOREIGN KEY constraints for parent references
- TR_Persons_Validate — AFTER INSERT, UPDATE trigger re-checks all axioms
- FN_HasAncestorCycle — scalar function walks the ancestor graph to detect cycles (Axiom 3)
- SP_Persons_Delete — prevents deletion of persons who are referenced as parents

**Security Measures**

| Threat | Mitigation |
| --- | --- |
| SQL Injection | All data access via **stored procedures** through **Dapper** with typed parameters — zero ad-hoc SQL |
| CSRF / Cross-Site Request Forgery | **Anti-forgery tokens** on every POST (enforced globally via AutoValidateAntiforgeryTokenAttribute) |
| XSS | Razor's default HTML-encoding; Content-Security-Policy header restricts script sources |
| Clickjacking | X-Frame-Options: DENY header |
| MIME sniffing | X-Content-Type-Options: nosniff header |
| HTTPS downgrade | UseHttpsRedirection() + HSTS (365 days, includeSubDomains) |
| Over-posting | [BindProperty] on a dedicated Input model — never bind Id from POST on edit |
| Integer overflow | short / SMALLINT for years; BIGINT for Id bounded to 0–65535 |

*Figure 5*: README file (second part)

The explicit constraints from Figure 10 correspond to the following 6 "business rules": no ruler may start or stop reigning over a country several times in the same day; no reign may end before starting; known start or end reigning day requires known corresponding month; known end reigning month requires known corresponding year. The 6 ones from Figure 11 correspond to: April, June, September, and November have only 30 days; February has at most 29; nobody may reign before being born or after death; no country may be simultaneously ruled by two persons, except for cases where at least one of them has sex 'N', or the two were married, or parent and child; nobody may reign twice in the same country in the same period; reign durations for males and females must be naturals at most equal to 140; any not ended reign must end in the year in which the corresponding ruler passed away (the □ symbol from this last constraint is the "*always in every future state*" quantifier from the temporal first-order logic).

Remarkably, Claude AI not only understood all explicit constraints as well, but it also clearly reverse their corresponding "business rules" in plain English.

At every such new step, Claude correspondingly updated the documentation shown in Figures 2 to 7. After each step, we tested the code and found errors that are presented in the Discussion Section. We also provided to Claude details on how to organize the GUI for accommodating the newly corresponding data management and query results forms and subforms.

For thorough testing, we manually imported with a T-SQL script over 3500 table lines from an older db: 1800 for *PERSONS*, 992 for *REIGNS*, 588 for *MARRIAGES*, 120 for *COUNTRIES*, and 47 for *TITLES*.

## Results

In total, the (E)MDM scheme that we provided to Claude has 26 functions, 33 explicit constraints, and 12 Datalog rules. From what Claude AI generated with our input, we accepted, 5 tables, 8 foreign keys, 13 unique ones, 18 check constraints, 4 triggers, 1 view, 29 stored procedures, and 1 assembly for the db, plus 5 forms and 8 subforms for the app. The app's source code consists of 46 files (24 of type razor, 13 cs, 5 sql, 2 json, 1 css, and 1 csproj), totaling 252KB (358 on disk). A .ps1 PowerShell script file stops the app's Windows server, (re)builds and (re)publishes the dotnet app, then (re)starts the Windows server. The build result consists of 1362 read-only files stored in 112 folders, occupying 352MB. The published Windows service has 365 read-only files in 12 folders, occupying 105MB.

Claude even added test data to tables, e.g., for the mythical Eve and Adam, as well as for their 3 generation of descendants.

```
Project Structure
PersonsApp/
├── Database/
│   └── Schema.sql          # Full DDL, triggers, stored procedures
├── Models/
│   └── Person.cs           # Domain model + ComputeAge()
├── Services/
│   ├── PersonValidator.cs  # All 8 axioms in C#
│   └── PersonRepository.cs # Dapper + stored procedures
├── Pages/
│   ├── Shared/
│   │   └── _Layout.cshtml  # Archival aesthetic layout
│   └── Persons/
│       ├── Index.cshtml(.cs)   # List all persons
│       ├── Create.cshtml(.cs)  # New person entry
│       ├── Edit.cshtml(.cs)    # Edit existing record
│       ├── Detail.cshtml(.cs)  # View + children list
│       └── _PersonForm.cshtml  # Shared form partial
├── Program.cs              # DI + security middleware
├── appsettings.json        # Connection string
└── PersonsApp.csproj       # .NET 8 project file
```

*Figure 6*: README file (third part)

Figure 13 shows the first app's page. Data on this page is displayed in ascending order of the persons' name and then birth year. Please note that in this screen capture, the first app submenu is open: from it you may choose between the web pages displaying the forms corresponding to the db 5 tables. This web page also has a filter for selecting the desired person's data, shown open in Figure 14, which narrows the search scope with every typed character, just like in the MS Access combo-boxes; its dropdown list is also presented in ascending order. Everywhere in this app where person data is referenced, like in this filter, they are displayed as *Name*, *Sex* (F, M, or N), which are mandatory, followed in parentheses by "b. birth year" comma "p. passed away year", if any (see the selected line from the dropdown list shown in Figure 14).

Figure 15 shows the *Countries* web page, while Figure 16 the *Titles* one. Both countries and titles are displayed in alphabetic order. Figure 17 shows the *Marriages* web page, which can be filtered by *Husband* or/and *Wife*; its data is sorted ascendingly by husband, marriage year, and wife. Figure 18 shows the *Reigns* web page, which can be filtered by *Country* or/and *Ruler*; its data is sorted ascendingly by country and from year (i.e., reign start) calendar data.

Figures 19 and 20 show all data for a person (selected in the dropdown list from Figure 14): the personal ones, the marriages, the children, and the reigns, if any. Figures 21 to 24 show the results of running the two queries defined in Figure 12 by the two Datalog programs.

**Setup**
1. **Create the database** — run Database/Schema.sql against SQL Server
2. **Configure connection** — edit appsettings.json (or use User Secrets):
3. "ConnectionStrings": { "PersonsDB":
   "Server=.;Database=PersonsDB;Trusted_Connection=True;TrustServerCertificate=True;"}
4. **Run** — dotnet run or open in Visual Studio

**Notes on the Formal Spec**
- **NAT(16)** is interpreted as a 16-bit natural number: $0 \ldots 2^{16}-1 = \mathbf{65\,535}$
- **Axiom 8b** allows posthumous birth: child may be born up to **1 year after** father's death
- **Axiom ③ (acyclic)** — the SQL FN_HasAncestorCycle CTE walks both Mother and Father edges; SQL Server's max recursion (100 by default) is a practical ceiling well beyond any realistic ancestry depth; the trigger sets MAXRECURSION 300 implicitly via the CTE structure
- Sex = 'N' (Neutral) persons are modelled as having no known biological parents by axiom ⑥, consistent with the formal specification
- **Age** is always recomputed from BirthYear and PassedAwayYear; it is never stored

*Figure 7*: README file (fourth part)

Mother • Name injective
Father • Name injective
Husband : MARRIAGES → PERSONS
Wife : MARRIAGES → PERSONS
MarriageYear : MARRIAGES → [-6500, CurrentYear()] ∪ NULLS
DivorceYear : MARRIAGES → [-6500, CurrentYear()] ∪ NULLS
Husband • Wife • MarriageYear injective
Husband • Wife • DivorceYear injective
(∀x∈MARRIAGES)(MarriageYear(x) ≤ DivorceYear(x))
(∀x∈MARRIAGES)(Sex(Wife(x)) = 'F')
(∀x∈MARRIAGES)(Sex(Husband(x)) = 'M')

*Figure 8*: Second step of specifications (first part)

The first 4 Datalog rules from Figure 12 compute the transitive closure of the set of *PERSONS* through their mothers and fathers, i.e., the set of all pairs of the type (*Ancestor*, *Descendant*). This set can be computed over the current instance of the *PERSONS* registry by clicking on the first option of the *Queries* submenu (*Persons Transitive Closure*). The result (see Figure 21) is presented in ascending order of the ancestors' name and birth year, then descendants' birth year and name.

The last 8 Datalog rules from Figure 12 compute the transitive closure of the set of *PERSONS* through their mothers and fathers for a desired person *seed*, i.e., the set of all pairs of the type (*Ancestor*, *seed*) merged with the one of type (*seed*, *Descendant*). This set can be computed over the current instance of the *PERSONS* registry by clicking on the second option of the *Queries* submenu (*Transitive Closure of Selected Person*). Figure 22 shows the web page for selecting the desired *seed* person.

$$(\forall x \in \text{MARRIAGES})(\text{MarriageYear}(x) \notin \text{NULLS} \Rightarrow ((\text{BirthYear}(\text{Husband}(x)) \in \text{NULLS} \vee \text{BirthYear}(\text{Husband}(x)) \leq \text{MarriageYear}(x)) \wedge (\text{PassedAwayYear}(\text{Husband}(x)) \in \text{NULLS} \vee \text{PassedAwayYear}(\text{Husband}(x)) \geq \text{MarriageYear}(x))) \wedge (\text{BirthYear}(\text{Wife}(x)) \in \text{NULLS} \vee \text{BirthYear}(\text{Wife}(x)) \leq \text{MarriageYear}(x)) \wedge (\text{PassedAwayYear}(\text{Wife}(x)) \in \text{NULLS} \vee \text{PassedAwayYear}(\text{Wife}(x)) \geq \text{MarriageYear}(x))))$$

$$(\forall x \in \text{MARRIAGES})(\text{DivorceYear}(x) \notin \text{NULLS} \Rightarrow ((\text{BirthYear}(\text{Husband}(x)) \in \text{NULLS} \vee \text{BirthYear}(\text{Husband}(x)) \leq \text{DivorceYear}(x)) \wedge (\text{PassedAwayYear}(\text{Husband}(x)) \in \text{NULLS} \vee \text{PassedAwayYear}(\text{Husband}(x)) \geq \text{DivorceYear}(x))) \wedge (\text{BirthYear}(\text{Wife}(x)) \in \text{NULLS} \vee \text{BirthYear}(\text{Wife}(x)) \leq \text{DivorceYear}(x)) \wedge (\text{PassedAwayYear}(\text{Wife}(x)) \in \text{NULLS} \vee \text{PassedAwayYear}(\text{Wife}(x)) \geq \text{DivorceYear}(x))))$$

$$(\forall x \in \text{MARRIAGES})(\text{Father}(\text{Husband}(x)) \neq \text{Father}(\text{Wife}(x)) \wedge \text{Mother}(\text{Husband}(x)) \neq \text{Mother}(\text{Wife}(x)) \wedge \text{Father}(\text{Wife}(x)) \neq \text{Husband}(x) \wedge \text{Mother}(\text{Husband}(x)) \neq \text{Wife}(x))$$

$$(\forall x \in \text{MARRIAGES})(\text{MarriageYear}(x) \notin \text{NULLS} \wedge \text{PassedAwayYear}(\text{Husband}(x)) \in \text{NULLS} \wedge \text{PassedAwayYear}(\text{Wife}(x)) \in \text{NULLS} \Rightarrow 0 \leq \text{isNull}(\text{DivorceYear}(y), \text{CurrentYear}()) - \text{MarriageYear}(x) \leq 140)$$

$$(\forall x,y \in \text{MARRIAGES})(x \neq y \wedge (\text{MarriageYear}(y) < \text{MarriageYear}(x) < \text{isNull}(\text{DivorceYear}(y), \text{CurrentYear}()) \vee \text{MarriageYear}(x) < \text{MarriageYear}(y) < \text{isNull}(\text{DivorceYear}(y), \text{CurrentYear}()) \vee \text{MarriageYear}(y) < \text{DivorceYear}(x) < \text{isNull}(\text{DivorceYear}(y), \text{CurrentYear}()) \vee \text{MarriageYear}(x) < \text{DivorceYear}(y) < \text{isNull}(\text{DivorceYear}(x), \text{CurrentYear}()))) \Rightarrow \text{Husband}(x) \neq \text{Husband}(y) \wedge \text{Wife}(x) \neq \text{Wife}(y))$$

$$(\forall x \in \text{MARRIAGES})(\text{BirthYear}(\text{Husband}(x)) \leq \text{BirthYear}(\text{Wife}(x)) \leq \text{isNull}(\text{PassedAwayYear}(\text{Husband}(x)), \text{CurrentYear}()) \vee \text{BirthYear}(\text{Wife}(x)) \leq \text{BirthYear}(\text{Husband}(x)) \leq \text{isNull}(\text{PassedAwayYear}(\text{Wife}(x)), \text{CurrentYear}()) \vee \text{BirthYear}(\text{Husband}(x)) \leq \text{PassedAwayYear}(\text{Wife}(x)) \leq \text{isNull}(\text{PassedAwayYear}(\text{Husband}(x)), \text{CurrentYear}()) \vee \text{BirthYear}(\text{Wife}(x)) \leq \text{PassedAwayYear}(\text{Husband}(x)) \leq \text{isNull}(\text{PassedAwayYear}(\text{Wife}(x)), \text{CurrentYear}()))$$

*Figure 9*: Second step of specifications (second part)

The result of computing such a transitive closure for a person (i.e., the corresponding genealogical trees) is presented in Figures 23 (the page's top) and 24 (the page's bottom). On one hand, this result omits from the one obtained with Datalog, in any pair, the *seed* person, as that person is understood and it does not make sense to repeat it on every line; on the other, this result is richer than the one obtained with pure Datalog (which is a logic language, with no algebraic provisions), as it assigns to each pair a *Generation* integer, as follows: the *seed* person has 0 (self); the parents –1; the grandparents –2; the great-grandparents –3, etc.; the children +1; the grandchildren +2, etc. The result is displayed in the ascending order of the *Generation* and then of persons' birth year and name.

Remarkably, all constraints are enforced both at the client level, in the .razor files, using embedded SQL in MS C#.NET, and at the SQL Server level, in the triggers, using the extended SQL MS T-SQL. Consequently, on one hand, some constraints may not even need access to the SQL Server (as all needed data is stored in the client's memory variables) or only need read-only access; on the other, constraints may not be violated not even by tampering with the db directly, circumventing the web client.

The web app generated is secure (see Figure 3):

- *SQL Injection*: zero ad-hoc SQL; all db calls go through named stored procedures via Dapper with typed parameters
- *CSRF*: anti-forgery tokens on every POST, enforced globally
- *XSS*: Razor's automatic HTML-encoding; strict Content-Security-Policy header
- *Clickjacking*: X-Frame-Options: DENY

- *HTTPS*: UseHttpsRedirection() + HSTS (365 days)
- *Over-posting*: [BindProperty] on a dedicated Input model; *x* id's are read-only on the edit forms.

$$\text{Country}: \text{COUNTRIES} \leftrightarrow \text{UNICODE}(128)$$
$$\text{CurrentCountry}: \text{COUNTRIES} \to \text{COUNTRIES} \cup \text{NULLS}, \text{acyclic}$$
$$\text{Title}: \text{TITLES} \leftrightarrow \text{UNICODE}(32)$$
$$\text{Ruler}: \text{REIGNS} \to \text{PERSONS}$$
$$\text{Country}: \text{REIGNS} \to \text{COUNTRIES}$$
$$\text{Title}: \text{REIGNS} \to \text{TITLES} \cup \text{NULLS}$$
$$\text{FromYear}: \text{REIGNS} \to [-6500, \text{CurrentYear}()]$$
$$\text{ToYear}: \text{REIGNS} \to [-6500, \text{CurrentYear}()] \cup \text{NULLS}$$
$$\text{FromMonth}: \text{REIGNS} \to [1, 12] \cup \text{NULLS}$$
$$\text{ToMonth}: \text{REIGNS} \to [1, 12] \cup \text{NULLS}$$
$$\text{FromDay}: \text{REIGNS} \to [1, 31] \cup \text{NULLS}$$
$$\text{ToDay}: \text{REIGNS} \to [1, 31] \cup \text{NULLS}$$
$$\text{FromDate} = \text{FromYear} \bullet \text{isNull}(\text{FromMonth}, 7) \bullet \text{isNull}(\text{FromDay}, 1)$$
$$\text{ToDate} = \text{isnull}(\text{ToYear}, \text{CurrentYear}()) \bullet \text{isNull}(\text{ToMonth}, 6) \bullet \text{isNull}(\text{ToDay}, 30)$$
$$\text{Ruler} \bullet \text{Country} \bullet \text{FromDate injective}$$
$$\text{Ruler} \bullet \text{Country} \bullet \text{ToDate injective}$$
$$(\forall x \in \text{REIGNS})(\text{ToDate}(x) \geq \text{FromDate}(x))$$
$$(\forall x \in \text{REIGNS})(\text{FromDay}(x) \notin \text{NULLS} \Rightarrow \text{FromMonth}(x) \notin \text{NULLS})$$
$$(\forall x \in \text{REIGNS})(\text{ToDay}(x) \notin \text{NULLS} \Rightarrow \text{ToMonth}(x) \notin \text{NULLS})$$
$$(\forall x \in \text{REIGNS})(\text{ToMonth}(x) \notin \text{NULLS} \Rightarrow \text{ToYear}(x) \notin \text{NULLS})$$

*Figure 10*: Third step of specifications (first part)

$(\forall x \in \text{REIGNS})((\text{FromMonth}(x) \in \{4,6,9,11\} \Rightarrow \text{FromDay}(x) \leq 30) \wedge (\text{ToMonth}(x) \in \{4,6,9,11\} \Rightarrow \text{ToDay}(x) \leq 30) \wedge (\text{FromMonth}(x) = 2 \Rightarrow \text{FromDay}(x) \leq 29) \wedge (\text{ToMonth}(x) = 2 \Rightarrow \text{ToDay}(x) \leq 29))$

$(\forall x \in \text{REIGNS})((\text{BirthYear}(\text{Ruler}(x)) \notin \text{NULLS} \Rightarrow \text{BirthYear}(\text{Ruler}(x)) \leq \text{FromYear}(x)) \wedge (\text{PassedAwayYear}(\text{Ruler}(x)) \notin \text{NULLS} \Rightarrow \text{ToYear}(x) \notin \text{NULLS} \wedge \text{PassedAwayYear}(\text{Ruler}(x)) \geq \text{ToYear}(x)))$

$(\forall x \in \text{REIGNS})((\text{BirthYear}(\text{Ruler}(x)) \notin \text{NULLS} \Rightarrow \text{BirthYear}(\text{Ruler}(x)) \leq \text{FromYear}(x)) \wedge (\text{PassedAwayYear}(\text{Ruler}(x)) \notin \text{NULLS} \Rightarrow \text{ToYear}(x) \notin \text{NULLS} \wedge \text{PassedAwayYear}(\text{Ruler}(x)) \geq \text{ToYear}(x))) (\forall x,y \in \text{REIGNS})(x \neq y \wedge \text{Country}(x) = \text{Country}(y) \wedge \text{Ruler}(x) \neq \text{Ruler}(y) \wedge (\text{FromDate}(x) \leq \text{FromDate}(y) < \text{ToDate}(x) \vee \text{FromDate}(y) \leq \text{FromDate}(x) < \text{ToDate}(y) \vee \text{FromDate}(x) < \text{ToDate}(y) \leq \text{ToDate}(x) \vee \text{FromDate}(y) < \text{ToDate}(x) \leq \text{ToDate}(y)) \Rightarrow \text{Sex}(x) = \text{‘N’} \vee \text{Sex}(y) = \text{‘N’} \vee (\text{Father}(\text{Ruler}(y)) = \text{Ruler}(x) \vee \text{Father}(\text{Ruler}(x)) = \text{Ruler}(y) \vee \text{Mother}(\text{Ruler}(y)) = \text{Ruler}(x) \vee \text{Mother}(\text{Ruler}(x)) = \text{Ruler}(y)) \vee (\exists z \in \text{MARRIAGES})(\text{Husband}(z) = \text{Ruler}(x) \wedge \text{Wife}(z) = \text{Ruler}(y) \vee \text{Husband}(z) = \text{Ruler}(y) \wedge \text{Wife}(z) = \text{Ruler}(x)))$

$(\forall x,y \in \text{REIGNS})(x \neq y \wedge \text{Country}(x) = \text{Country}(y) \wedge (\text{FromDate}(x) <= \text{FromDate}(y) < \text{ToDate}(x) \vee \text{FromDate}(y) \leq \text{FromDate}(x) < \text{ToDate}(y) \vee \text{FromDate}(x) < \text{ToDate}(y) \leq \text{ToDate}(x) \vee \text{FromDate}(y) < \text{ToDate}(x) \leq \text{ToDate}(y)) \Rightarrow \text{Ruler}(x) \neq \text{Ruler}(y))$

$(\forall x \in \text{REIGNS})(\text{Sex}(\text{Ruler}(x)) \neq \text{‘N’} \Rightarrow \text{isNull}(\text{ToYear}(y), \text{CurrentYear}()) - \text{FromYear}(x) \leq 140)$

$(\forall x \in \text{PERSONS})(\forall y \in \text{REIGNS})(\text{PassedAwayYear}(x) \notin \text{NULLS} \wedge \text{Ruler}(y) = x \wedge \text{ToYear}(y) \in \text{NULLS} \Rightarrow \square(\text{ToYear}(y) = \text{PassedAwayYear}(x)))$

*Figure 11*: Third step of specifications (second part)

- *DB connection string*: accessed through a .json file that uses the "Server=." and "Trusted_Connection = True" syntax, which means that the name of the SQL Server is, implicitly, the one of the local one, while the credentials of its authorized user are implicitly those of your own Windows user, which already has access, since you just ran the schema script as that user (i.e., no username or password needed).

$$\text{TransClosure(Ancestor, Descendant)} \leftarrow \text{PERSONS}(x=\text{Descendant}, \text{Father}=\text{Ancestor})$$
$$\text{TransClosure(Ancestor, Descendant)} \leftarrow \text{PERSONS}(x=\text{Descendant}, \text{Mother}=\text{Ancestor})$$
$$\text{TransClosure(Ancestor, Descendant)} \leftarrow \text{TransClosure}(x, \text{Descendant}), \text{PERSONS}(x, \text{Father}=\text{Ancestor})$$
$$\text{TransClosure(Ancestor, Descendant)} \leftarrow \text{TransClosure}(x, \text{Descendant}), \text{PERSONS}(x, \text{Mother}=\text{Ancestor})$$
$$\text{kPersTransClosure(Ancestor, k)} \leftarrow \text{PERSONS}(x=k, \text{Father}=\text{Ancestor})$$
$$\text{kPersTransClosure(Ancestor, k)} \leftarrow \text{PERSONS}(x=k, \text{Mother}=\text{Ancestor})$$
$$\text{kPersTransClosure(Ancestor, k)} \leftarrow \text{kPersTransClosure}(x, k), \text{PERSONS}(x, \text{Father}=\text{Ancestor})$$
$$\text{kPersTransClosure(Ancestor, k)} \leftarrow \text{kPersTransClosure}(x, k), \text{PERSONS}(x, \text{Mother}=\text{Ancestor})$$
$$\text{kPersTransClosure(k, Descendant)} \leftarrow \text{PERSONS}(x=\text{Descendant}, \text{Father}=k)$$
$$\text{kPersTransClosure(k, Descendant)} \leftarrow \text{PERSONS}(x=\text{Descendant}, \text{Mother}=k)$$
$$\text{kPersTransClosure(k, Descendant)} \leftarrow \text{kPersTransClosure}(x, \text{Descendant}), \text{PERSONS}(x=\text{Descendant}, \text{Father}=k)$$
$$\text{kPersTransClosure(k, Descendant)} \leftarrow \text{kPersTransClosure}(x, \text{Descendant}), \text{PERSONS}(x=\text{Descendant}, \text{Mother}=k)$$

*Figure 12*: Fourth step of specifications

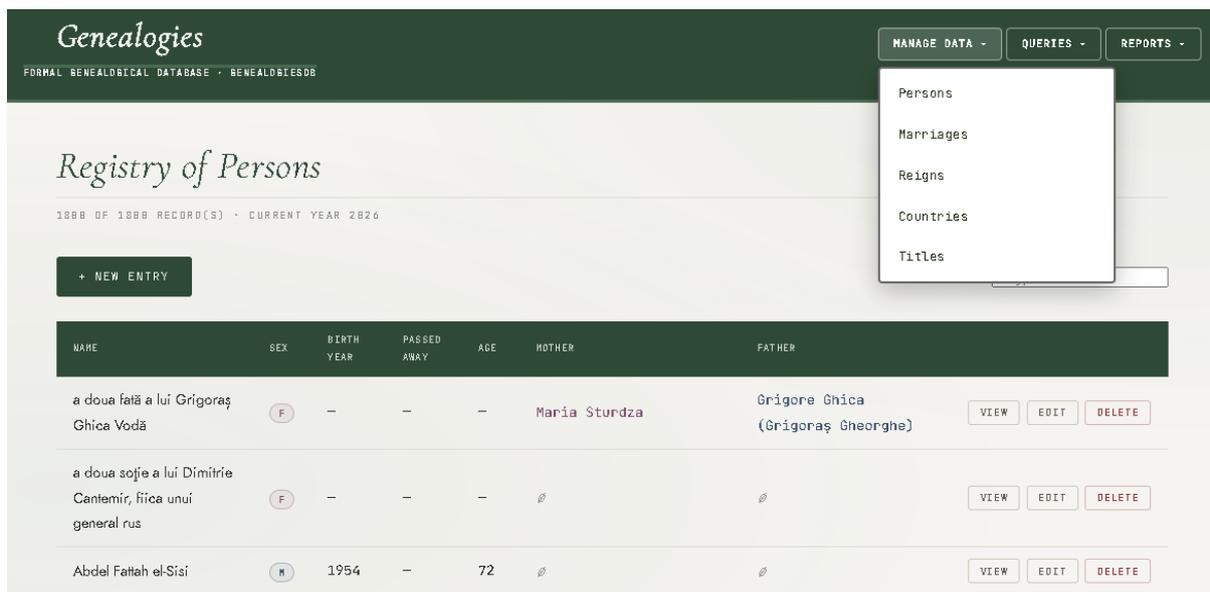

*Figure 13*: Top of the *Genealogies* web app's first page with the first submenu open

# Discussion

To finish this web app, we worked some 40h each. 270 A4 pages of dialog with Claude AI were necessary. In total, 205 rebuilds and republishes were done. Nearing the end of it, Claude AI displayed a message informing us that we reached 75% of the free resources for a week and invited us to subscribe to its Opus version.

We never used in our project specifications submitted to Claude AI the (E)MDM abbreviations. For example, the last constraint from Figure 10 is of type existence (see [31]) and is simply written in (E)MDM as *ToMonth |— ToYear*.

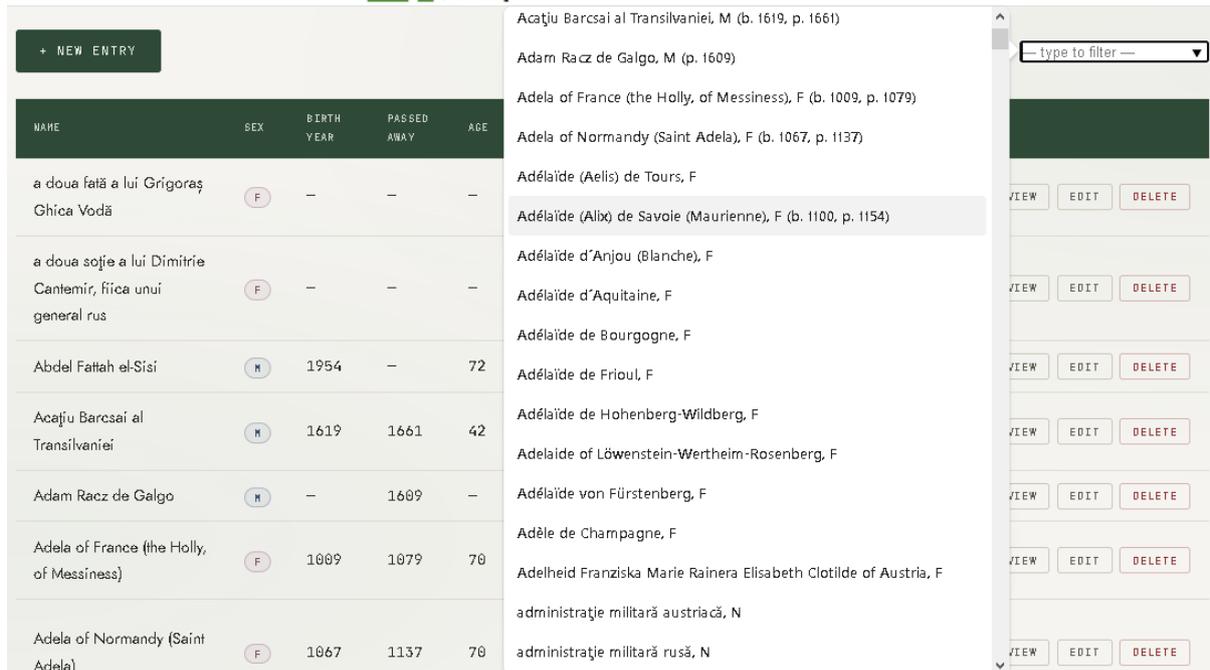

*Figure 14*: Top of the *Genealogies* web app's first page with the filter dropdown open

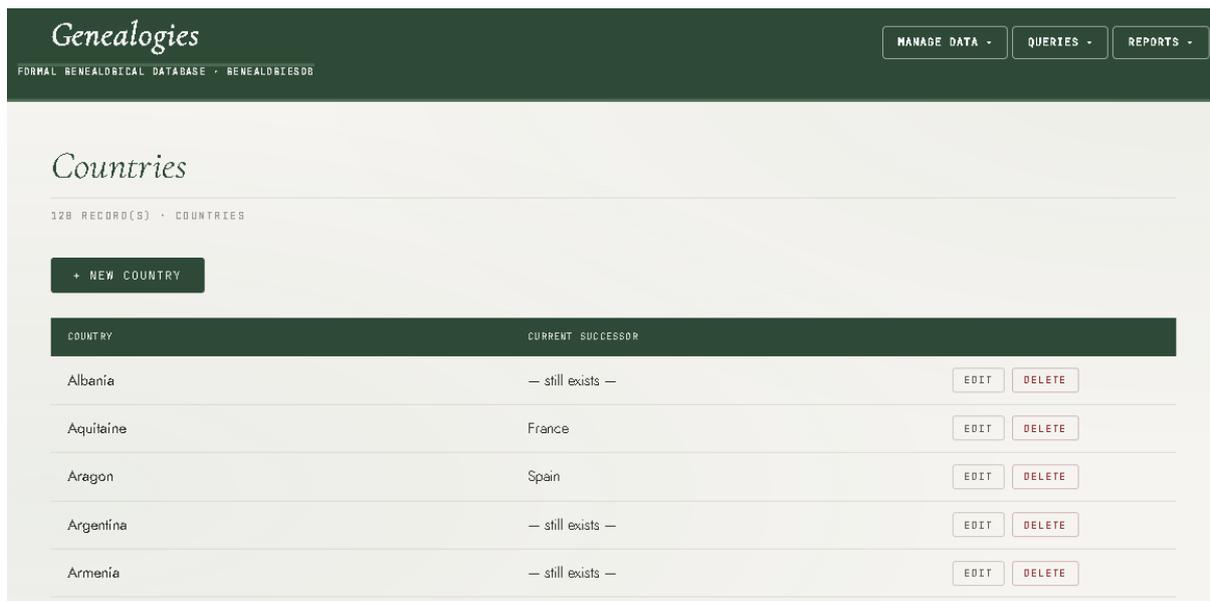

*Figure 15*: Top of *Countries* page of the *Genealogies* web app

We did not like the first GUI black background initially generated by Claude, which remembered the ancient DOS/Unix/Linux world. From the 4-5 alternatives we were offered, we chose the "botanic" dark green one, to match it with the color of the proposed 🌳 app favicon (probably inspired by the genealogical trees stored and computed). Generally, we did not pay much attention to the graphic: correct app functioning, constraint enforcement, and query computation were our main goals. We only added some italicizations, bolding, copyright notice, and rephrased some comments and messages to make them as clear as possible, although font sizes should also be enlarged for almost everything, except the app name (*Genealogies*).

*Figure 16*: Top of *Titles* page of the *Genealogies* web app

*Figure 17*: Top of *Marriages* page of the *Genealogies* web app

Conceptually, the generated GUI is rather childish, full of buttons offering clumsy functionality; e.g., you cannot delete several lines from a datasheet as a block, but only one by one; to add a new entry or update an existing one you need to press a button and work in another web page, then press other buttons to save your work and get back to the datasheet form. Unfortunately, the professional simplicity and elegance of, e.g., MS Access forms is history: there, you can select any number of rows, not necessarily contiguous, and ask for their deletion only by pressing the Delete key, you can add a new record on the blank line always available at the datasheet bottom, and you can update any record directly in the datasheet.

Two GUI related incidents amazed us. First, after providing the subschema from Figure 1, the app's menu was horizontal, with two options, *All Records* and *New Entry*; after adding the subschema from Figures 8 and 9, buttons

*Marriages*, *New Marriage*, *Countries*, and *Titles* were concatenated (with the latter two already outside the page screen width); after adding the subschema from Figures 10 and 11, we asked Claude to reorganize the menu as it is now, i.e., with only 3 options, each one opening vertical submenus (see Figure 14); it took Claude more than 2h to do it, (re)modifying the corresponding razor file more than a dozen times!!

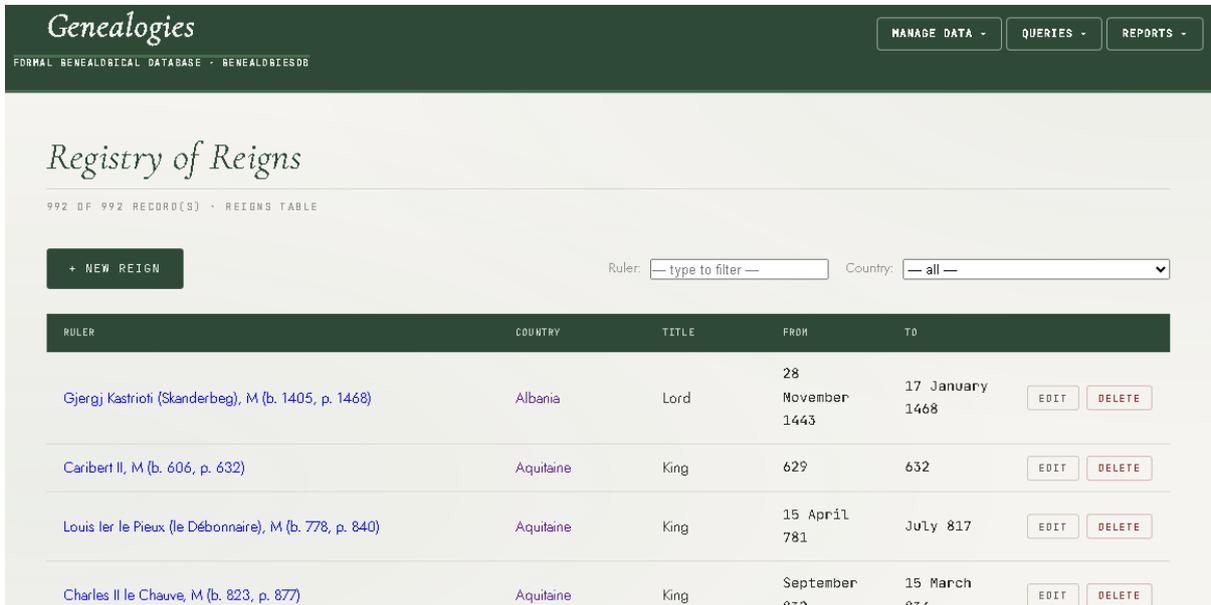

*Figure 18*: Top of *Reigns* page of the *Genealogies* web app

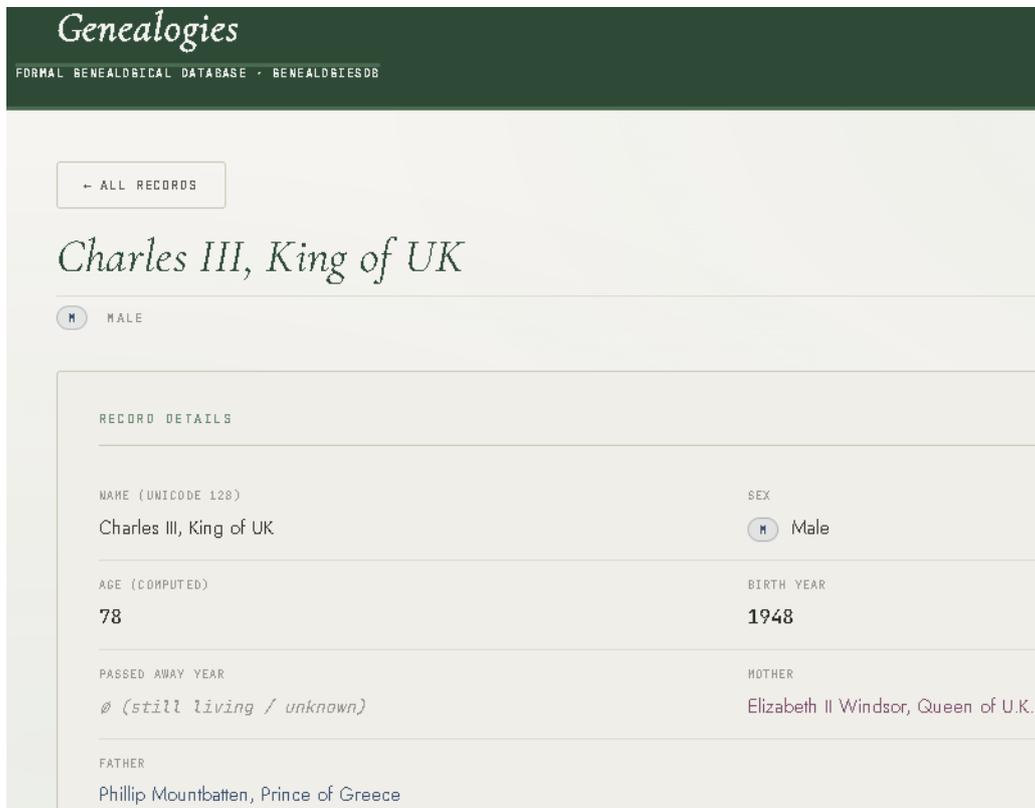

*Figure 19*: Top of a *Person* page of the *Genealogies* web app

*Figure 20*: Bottom of a *Person* page of the *Genealogies* web app

*Figure 21*: Top of *Persons Transitive Closure* page of the *Genealogies* web app

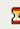

**Figure 22**: *Genealogies* web app page for selecting desired person for whom to compute the *transitive closure*

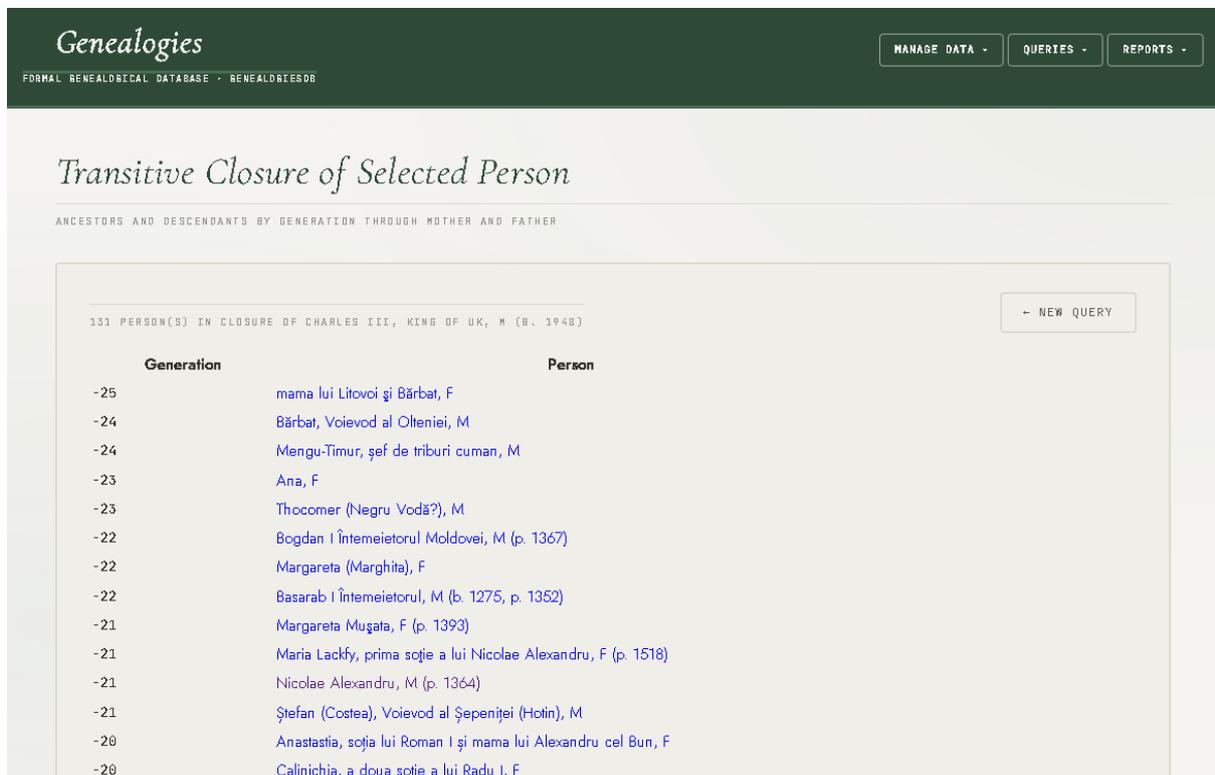

**Figure 23**: Top of a *Transitive closure of Selected Person* page of the *Genealogies* web app

```
-6                        Eduard Augustus, Duce de Kent și Strathearn, M (b. 1767, p. 1820)
-6                        Laszlo Rhedey de Kis-Rhede, M (b. 1775, p. 1835)
-6                        Agnes Inczady de Nagy-Varad, F (b. 1788, p. 1856)
-5                        Albert, Prinț de Saxa-Coburg și Gotha, M
-5                        Augusta Wilhelmina Louisa of Hesse-Kassel, F
-5                        Adolphus Frederick, Duke of Cambridge, M (b. 1774, p. 1850)
-5                        Alexander Paul of Wurttemberg, M (b. 1804, p. 1885)
-5                        Claudine Susanna Rhedey de Kis-Rhede, F (b. 1812, p. 1841)
-5                        Victoria, Regină a Regatului Unit, F (b. 1819, p. 1901)
-4                        Alexandra of Denmark, F
-4                        Mary Adelaide Wilhelmina Elizabeth of Cambridge, F (b. 1833, p. 1897)
-4                        Francis, Duke of Teck (Paul Charles Louis Alexander von Hohenstein), M (b. 1837, p. 1900)
-4                        Edward VII Albert of Great Britain, M (b. 1841, p. 1910)
-3 (great-grandparent)    George V Frederick Ernest Albert of Great Britain, M (b. 1865, p. 1936)
-3 (great-grandparent)    Mary of Teck, F (b. 1867, p. 1953)
-2 (grandparent)          Elizabeth Angela Marguerite Bowes-Lyon, The Queen Mother, F
-2 (grandparent)          George VI Albert Frederick Arthur og Great Britain, M (b. 1895, p. 1952)
-1 (parent)               Phillip Mountbatten, Prince of Greece, M (b. 1921, p. 2021)
-1 (parent)               Elizabeth II Windsor, Queen of U.K., F (b. 1926, p. 2022)
 0 (self)                 Charles III, King of UK, M (b. 1948)
+1 (child)                William, Prince of Wales, M (b. 1982)
+1 (child)                Harry, Duke of Sussex, M (b. 1984)
+2 (grandchild)           George of Wales, M (b. 2013)
+2 (grandchild)           Charlotte of Wales, F (b. 2015)
+2 (grandchild)           Louis of Wales, M (b. 2018)
+2 (grandchild)           Archie Harrison Mountbatten-Windsor, M (b. 2019)
+2 (grandchild)           Lilibet Diana „Lili" Mountbatten-Windsor, F (b. 2021)
```

*Figure 24*: Bottom of a *Transitive closure of Selected Person* page of the *Genealogies* web app

Then, for the web page shown in Figure 22, Claude initially generated a razor page only containing a picker identical to the one from Figure 14, trying to implement the simplest strategy possible: let the user pick the desired person from the dropdown list, narrowing the search scope with every key stroke, and when he/she presses Enter fire the corresponding SQL query and navigate to the web page shown in Figure 23 for displaying its results.

In total, it tried for almost 10h to implement it, with dozens and dozens of trials and errors in vain. Not even the *Compute* button from Figure 22 that we proposed was fully helpful. In the end, Claude declared itself unable to program this, and replaced the corresponding razor page with a HTML one!!! This is why the picker from Figure 22 is the only app's one in which narrowing the search scope is done only after the first pressed key!

There were several other implementation incidents related to MS Blazor, whose core idea is elegant otherwise: write C# instead of JavaScript and let the framework handle the browser/server communications via SignalR WebSockets. Unfortunately, its implementation has at least the following 4 serious issues:

*Figure 25*: Bottom of the *Person\create* page of the *Genealogies* web app

1. DOM synchronization is fragile: Blazor's virtual DOM diffing breaks in subtle ways when dealing with large datasets, <select> elements, and components that re-render frequently. In our case, a <select> with 1800 persons silently breaks all button click handlers on the page shown in Figure 22, without any error message or, at least, a log entry. This is a serious design flaw.
2. The event handling model is unreliable: @onclick on buttons in layout components doesn't work; async Task onclick handlers sometimes don't trigger re-renders; @bind on selects loses state. And these are not edge cases: they are the most basic UI interactions imaginable!
3. The abstraction leaks constantly: you need to know all the time about SignalR message sizes, DOM diffing algorithms, render cycles, StateHasChanged, OnAfterRenderAsync, all being supposed to be hidden from the developer!
4. The debugging experience is poor: when something doesn't work, there is often no error — just silence. We spent long hours trying to make work one of the simplest web pages imaginable (the one from Figure 22), not because the problem was complex, but because we had no signal to follow.

Again, comparison with MS VBA is insightful: a single VBA ComboBox1.ListIndex = -1 statement versus hours of @bind, @onchange, StateHasChanged, string-backed properties, JS interop, and finally a minimal application programming interface (API) endpoint written in C# to render plain HTML — just to get a dropdown list that works…

To finish our discussion about the apps GUI on a bright note, we were pleasantly surprised that, on every /create page (i.e., for adding new records to the five db tables), Claude AI displayed at their bottom, under the heading "ACTIVE AXIOMS", the set of constraints that are enforced for that set, which we did not asked for, but consider an excellent initiative (see, e.g., Figure 25).

The dialogue with Claude AI went extremely smoothly: only once did it not understand our request (out of hundreds), so it asked for details. Generally, we recommend asking fewer things in every request: e.g., Claude found extremely rich our specs from Figures 8 and 9, as well as those from Figures 10 and 11.

Not only for them, at least a dozen of times during that work week, Claude stopped in the middle of an answer and displayed the message "Maximum workload per round reached!"; fortunately, if you click on the *Continue* button that accompanies it, Claude resumes the dialog.

We would also recommend not to send complex, not even medium complex screen shots: it is better to explain what happened or what you want in plain English, otherwise you risk longer attempts to get an answer. Moreover, be aware that if Claude cannot answer after 10 attempts, it aborts the question and reinstates your query; if you experience a second abortion after another 10 attempts, do not retry the same question / screen shot, as it is in vain: rephrase your question, simplifying it.

Conceptually, Claude AI is a very intelligent co-worker, but still "human", making reasoning mistakes from time to time. For example, in our second round of specs (triggered by the subschema from Figures 8 and 9) it added the superfluous tuple constraint *Wife ≠ Husband*, which is implied by $Im(Sex \circ Husband)$ = 'M' and $Im(Sex \circ Wife)$ = 'F'. Even less understandable were his questions related to excluding incestuous marriages from Figure 9, on whether adding two more constraints to it, namely *Father ° Husband ≠ Wife* and *Mother ° Wife ≠ Husband*, although, trivially, they are not needed, being implied by the same two constraints as above.

Much more severe was the fact that Claude, repeatedly, did not fully, but only partially enforce the constraints that involve several sets or several elements of the same set. For example, all constraints from Figure 9 except for the last but one one, involve not only functions defined on *MARRIAGES*, but also functions defined on *PERSONS*; however, Claude Code initially enforced all of them only on the columns of the *MARRIAGES* table; similarly, but more subtly, the last but one constraint, which involves two distinct elements of the same set, must be symmetrically enforced for both, just as specified in our constraint; however, Claude Code ignored half of it and only enforced it in one direction. Fortunately, when we draw its attention to it, Claude immediately added the code needed to fully enforce all of them for the columns of table *PERSONS* too, as well as on the opposite direction for the last but one one.

Unfortunately, the same situation was repeated for the constraints from Figure 11. What was the worst about it is the fact that not only when correcting Blazor Server's unexpected behaviors, but also when updating the SQL triggers code, the new or updated files that Claude provided us with were lacking the code it previously injected for dually enforcing all constraints from Figures 9 to 11! This is when we started to add best practice rules (that Claude AI calls *meta-axioms*) in our dialogue, topic that we will deal with at the end of this section.

Another conceptual mistake that Claude Code committed was to silently delete valid stored data for accepting as valid a value that, otherwise, would violate a constraint: e.g., in one of our tests we successfully saved data on a daughter of a woman born in 2021 whose birth year was unknown; then, we modified her birth year to 1000; when we clicked the save button, instead of rejecting this unplausible value, the app silently deleted the data of her mother and saved the update!! Of course, this triggered another meta-axiom (see the end of this Section).

Another conceptual error that triggered the necessity of adding a new meta-axiom manifested in the initial versions of data filters: e.g., after selecting as husband (see Figure 17) a male with known birth year, you were then presented in the adjacent dropdown filter list of wives with data of all female persons, including the mythical Eve.

Although not that frequently, there were also SQL conceptual and coding errors. For example, the constraint formalizing the fact that males and females may live for at most 140 years (see Figure 1) was first enforced only for living persons (i.e., those with unknown passed away year). Much more severe, the initial Common Table Expressions (CTE) recursive SQL statements for both enforcing the *Mother • Father* acyclic constraint (see Figure 1) and computing the transitive closures for a selected person (see Figure 23) were incorrect: none of them detected cycles having length greater than one. Fortunately, they were promptly corrected after we signaled them.

Less severe, was the fact that, even after establishing the meta-axiom on surrogate table primary keys, Claude generated for them the name "id" instead of "x" for both *COUNTRIES*, *MARIAGES*, and *REIGNS*.

Severe indeed was generating for *TITLES* the primary key *Title* (see Figure 10) and, especially, silently replacing "<" with "≤" in the implementation of the third constraint from Figure 11, which caused the impossibility to store two adjacent reigns in the same country when the second started in the same year in which the first one ended.

Finally, there were a few times when Claude Code recognized that it fell in the trap of ignoring known technology issues. For example, once it ignored that Dapper's ExecuteAsync has a known issue with swallowing trigger rollbacks in some configurations, while SqlCommand does not. On another occasion, it initially ignored that @bind has a known issue where it does not always re-render the selected option correctly after a programmatic value change, while value + @onchange + explicit selected attributes give the browser unambiguous instructions about which option is selected at every render.

Boris Cherny, Claude Code's creator, emphasized repeatedly in his podcast that there isn't a single correct way to use his product, which was "*intentionally built in a way that you can use it, customize it, and hack it however you like.*" The mechanism to "hack" it is adding, explicitly, in its CLAUDE.md project files, or implicitly meta-axioms (*rules*).

Basically, CLAUDE.md files are project root markdown files that Claude Code reads at the beginning of every session: they are equivalent to explaining the entire project workflow to a new developer joining your team. We did not do it explicitly but let Claude Code do it implicitly: the difference is that, implicitly, sometimes you must remind it a rule, when you see that it has not automatically stored it in the corresponding .md file.

Here is the set of rules (meta-axioms) that we established for this project with Claude:

1. Never rename the concepts from the specs (i.e., set and function names, in our case).
2. Even if the corresponding function is absent from the specs, every fundamental db table must a have a surrogate primary key called "*x*".
3. Never declare primary keys that are not surrogates (i.e., surrogate integer primary keys are always preferable to natural ones, because they are stable, compact, and immune to data changes).
4. Any foreign key must reference a primary surrogate key, never a natural one.
5. Never delete / omit valid code when replacing a file with a new version.
6. Never fall into the traps of known technology issues.
7. Never reject a record that could become valid upon replacing NULLs with legitimate non-NULL values.
8. A record should be rejected if and only if every possible replacement of its NULL values with valid non-NULL values would violate at least one axiom.
9. An application may only accept or reject data submitted by the user and must never modify stored data without explicit user consent.
10. Never offer users the chance to choose from invalid data values (that the app will then have to reject).
11. Any constraint (axiom) may be violated by changing the values of any of its functions involved.
12. Enforcing a constraint that involves two distinct elements of the same set must be done in both directions.
13. Superfluous constraints (axioms) should be never enforced.
14. The undirected graph of a product of acyclic self-maps is not necessarily acyclic.

## Conclusion

We successfully built using Claude AI a safe web application, starting only with a(n) (E)MDM schema. The main contribution of this paper is proving that Claude AI almost perfectly understands mathematical function definitions, closed first-order predicate logic with equality formulas, and even Datalog logic inference rules. This is also the most eloquent proof of the value, power, and eventual bright future of our (Elementary) Mathematical Data Model. This is also proof that we do not need any other tools for modeling as programming: it is sufficient that data related modelers, as well as db and software architects review their K-12 and first college year math knowledge. The dream of many people, including the corresponding author of this paper, finally came true: we can now speak mathematics to computers.

We also provided a set of 14 best practice rules that are meaningful for any other db software application project.

The experience we report in this paper also proves that Claude AI is an excellent co-worker as a developer, but it still makes mistakes, both of conceptual and programming types, so that it cannot yet completely replace developers. Probably, its Opus, not freely available version, is better but we doubt that it is perfect. Dually, even its Sonnet free version that we used is significantly improving productivity.

Generally, most of Claude's coding errors were not conceptual, but of programming nature and, especially, due to the MS Blazor Server, in the .razor files' C# one, while much less frequent in the SQL triggers and stored procedures.

It seems that the MS Blazor Server is genuinely not designed for pages with thousands of DOM elements, but only for business forms, not large data grids or huge dropdowns. We recommend using Blazor technology only for complex stateful forms with server-side validation (e.g., our *Persons*, *Marriages*, and *Reigns* edit pages). For simple read-only interactions (e.g., the one in Figure 22), plain HTML/minimal API is preferable, by far.

We are convinced that developers who will not rapidly adopt meaningfully using AI tools will soon have to switch to other types of jobs. Dually, it would be a huge mistake to hire non-developers for developer positions, as the dialogue with Claude AI not only uses specific jargon, but, e.g., often, Claude does not offer to download full new file versions, but only statements that you must either add to specific places or replace specific code in existing files.

We are extremely confident that Claude AI, already impressively powerful, will continue to accelerate its growth and polish itself, thus forcing its concurrence to become better as well.

## Conflict of interest

The authors declare that the research was conducted in the absence of any commercial or financial relationships that could be construed as a potential conflict of interest.

## Acknowledgements

This research was not sponsored by anybody and nobody other than its authors contributed to it. The corresponding author always recalls with pleasure the contributions made by some of his outstanding former students: Lavinia Crasovschi for the (E)MDM, Adrian Mocanu and Sabina-Maria Motoc for *MatBase*.

## References

1. Mancas C. "The (Elementary) Mathematical Data Model revisited". PriMera Scientific Engineering 5.4 (2024): 78-91.
2. Mancas C. "*MatBase* – A Tool for Transparent Programming while Modeling Data at Conceptual Levels". In: Proc. 5th Int. Conf. on Comp. Sci. & Inf. Techn. (CSITEC 2019), AIRCC Pub. Corp. Chennai, India (2019): 15-27.
3. Abiteboul S, Hull R and Vianu V. "Foundations of Databases". Addison-Wesley, Reading, MA (1995).
4. Codd EF. "A relational model for large shared data banks". CACM 13.6 (1970): 377-387.
5. Mancas C. "Conceptual Data Modeling and Database Design: A Completely Algorithmic Approach. Volume 1: The Shortest Advis able Path". Apple Academic Press, Waretown, NJ (2015).
6. Chen PP. "The entity-relationship model. Toward a unified view of data". ACM TODS 1.1 (1976): 9-36.
7. Thalheim B. "Entity-Relationship Modeling: Foundations of Database Technology". Springer-Verlag, Berlin, Germany (2000).
8. Mancas C. "*MatBase* Metadata Catalog Management". Acta Scientific Computer Sciences 2.4 (2020): 25–29.

9. Mancas C. "On Modelware as the 5th Generation of Programming Languages". Acta Scientific Computer Sciences 2.9 (2020): 24–26.
10. Foote K. D. "A Brief History of Large Language Models" (2023). https://www.dataversity.net/articles/a-brief-history-of-large-language-models/
11. Microsoft. GitHub (2026). https://github.com/features/copilot
12. Anysphere. Cursor (2026). https://cursor.com/
13. Cognition AI. Windsurf ai (2026). https://windsurf.com/editor
14. Replit. Replit AI (2026). https://replit.com/
15. Anthropic Academy. Build with Claude (2026). https://www.anthropic.com/learn/build-with-claude
16. Claude Code Docs. Claude Code Overview (2026). https://code.claude.com/docs/en/overview
17. Claude Code Docs. How Claude Code Works (2026). https://code.claude.com/docs/en/how-claude-code-works
18. Faisal M. I set up Claude Code the way its creator does, and the difference is night and day (2026). https://www.xda-developers.com/set-up-claude-code-like-boris-cherny/?fbclid=IwY2xjawQnfkVleHRuA2FlbQIxMQBzcnRjBmFwcF9pZBAyMjIwMzkxNzg4MjAwODkyAAEer8Yn1EDYa6CQLTQuOD4i8vHp-bMMjGCR46JqN7_0zRWKt4daSuXgX1y13QM_aem_uFz01GtKUrsf2vqmO0jChw
19. Haynes R. "The Secure Path to AI-Powered Development A Strategy for Speed, Security, and Trust." O'Reilly Media, Inc., Santa Rosa, CA.
20. Moluck E. Claude Code and What Comes Next (2026). https://www.oneusefulthing.org/p/claude-code-and-what-comes-next
21. Sewell S. How I use Claude Code (+ my best tips) (2025). https://www.builder.io/blog/claude-code interface
22. Vannoy D. Claude Code Essentials for Data Professionals (2026). https://dustinvannoy.com/2026/01/08/claude-code-essentials/
23. Dutt A. Claude Code: A Guide With Practical Examples (2026). https://www.datacamp.com/tutorial/claude-code
24. Unicodedeveloper. 10 Must-Have Skills for Claude (and Any Coding Agent) in 2026 (2026). https://medium.com/@unicodeveloper/10-must-have-skills-for-claude-and-any-coding-agent-in-2026-b5451b013051
25. OpenAI. ChatGPT (2026). https://chatgpt.com/
26. Team Descope. Developer's Guide to AI Coding Tools: Claude vs. ChatGPT (2026). https://www.descope.com/blog/post/claude-vs-chatgpt
27. Khan U. A. Is Claude Code About to Disrupt the Billion-Dollar AI Coding Tools Market? (2025). https://medium.com/data-science-collective/is-claude-code-about-to-disrupt-the-billion-dollar-ai-coding-tools-market-813a67b8a5f0
28. TOP 5 WEBSITE BUILDERS. Best AI App Builders (2026) https://top5-websitebuilders.com/best-ai-app-builders/?network=g&gad_source=1&gad_campaignid=23531990068&gbraid=0AAAAADix23d-FjY26nGCWNp-xsdJlL826&gclid=Cj0KCQjw9-PNBhDfARIsABHN6-1-G5vi83x-0tKPd8kGRr5EBdy-crDqrrc4exsnMNY-UKYeQxQgtJ4aAoYIEALw_wcB
29. Ferri A. Claude Code saved us 97% of the work — then failed utterly (2025). https://www.thoughtworks.com/insights/blog/generative-ai/claude-code-codeconcise-experiment
30. The Pragmatic Engineer. Building Claude Code with Boris Cherny (2026). https://www.youtube.com/watch?v=julbw1JuAz0
31. Mancas C. "On Enforcing Existence and Non-Existence Constraints in *MatBase*". PriMera Scientific Engineering 4.6 (2024): 04-12.